# On the dipole moment of quantized vortices generated by flows

S. I. Shevchenko[1], A. M. Konstantinov[2]

[1] B.Verkin Institute for Low Temperature Physics and Engineering of the National Academy of Sciences of Ukraine, 47 Lenin Ave., Kharkov 61103, Ukraine
[2] V. N. Karazin Kharkov National University, 4 Svobody Sq., Kharkov 61022, Ukraine

Abstract

The polarization charge $\rho$ of an inhomogeneous superfluid system is expressed as a function of the order parameter $\Phi(\mathbf{r}_1, \mathbf{r}_2)$. It is shown that if the order parameter changes on macroscopic distances, the polarization charge $\rho_{pol}$ is proportional to $A\nabla^2 n$, and the polarization $\mathbf{P}$ is proportional to $A\nabla n$, where $n$ is the density of the system. For noninteracting atoms the proportionality coefficient $A$ is independent of density, and in the presence of interaction $A$ is proportional to $n$. The change of the Bose gas density is found in the presence of a flow $\mathbf{w} = \mathbf{v}_n - \mathbf{v}_s$ passing the vortex. It is found that a vortex in a superfluid film creates an electric potential above the film. This potential has the form of a potential of a dipole, allowing to assign a dipole moment to the vortex. The dipole moment is a sum of two terms, the first one is proportional to the relative flow velocity $\mathbf{w}$ and the second one is proportional to $[\boldsymbol{\kappa} \times \mathbf{w}]$, where $\boldsymbol{\kappa}$ is the vortex circulation.

Experiments [1, 2] carried out at ILT in Kharkov last 10 years revealed that a standing wave of second sound in He II was accompanied by appearance of electric potential of the order $10^{-7}$ V in the cell filled with helium. These experiments attracted great interest and stimulated theoretical investigations where attempts were made to explain the observed phenomena. Although an explanation of the experiments has not been found, several interesting effects associated with polarization of normal and superfluid dielectric systems were predicted, in particular, polarization of any dielectric system due to its accelerated motion [3], a possibility of a flexoelectric effect in quantum liquids and gases (polarization caused by system's inhomogeneity even without interaction between the atoms) [3, 4]. Using the results of [3], the author of [5] showed that vortex motion of atoms in a quantum liquid leads to their polarization (an axially symmetric polarization "hedgehog" appears around the vortex). In the presence of a magnetic field the core of a quantized vortex in a superfluid system obtains a polarization electric charge [6, 7]. The compensating charge of the opposite sign appears on the surface of the system and is generally at a macroscopically large distance from the vortex core. In [8] an idea of electric polarization of vortices by an external flow passing them has been stated. This article is devoted to development of this idea.

The consideration is performed in a model of a diluted Bose gas whose particles have internal degrees of freedom associated with motion of electrons relative to the nucleus [9]. It is supposed that each atom consists of two fermions, the valence electron with mass $m_e$ and the atomic "core" with mass $m_c$, the motion of electrons inside the core is neglected. $\hat{\Psi}_e^+(\mathbf{r})$ and $\hat{\Psi}_c^+(\mathbf{r})$ are creation operators of an electron and an atomic core respectively at the point $\mathbf{r}$ which satisfy Fermi commutation relations. As far as the electron and the atomic core form a bound state, we can expect that the wave function of the system has the form of a generalized coherent state [10, 6, 7]

$$|\Phi_0\rangle = \exp\left\{\int d\mathbf{r}_1 d\mathbf{r}_2 \left[\hat{\Psi}_c^+(\mathbf{r}_1)\Phi(\mathbf{r}_1,\mathbf{r}_2)\hat{\Psi}_e^+(\mathbf{r}_2) - h.c.\right]\right\}|0\rangle. \qquad (1)$$

The complex function $\Phi(\mathbf{r}_1, \mathbf{r}_2)$ (order parameter) present in this equation is found from the condition of minimum of the energy $E$. After variating the difference $E - \mu N$ ($\mu$ is the chemical potential and $N$ is the number of particles in the system) by $\Phi(\mathbf{r}_1, \mathbf{r}_2)$ and equating the result to zero an equation was obtained [9]

$$\left( -\frac{\hbar^2}{2m_c}\frac{\partial^2}{\partial \mathbf{r}_1^2} - \frac{\hbar^2}{2m_e}\frac{\partial^2}{\partial \mathbf{r}_2^2} - U_{ce}(|\mathbf{r}_1 - \mathbf{r}_2|)\right)\Phi(\mathbf{r}_1,\mathbf{r}_2) +$$

$$+ \int R_d(\mathbf{r}_1,\mathbf{r}_2,\mathbf{r}_3,\mathbf{r}_4)\Phi(\mathbf{r}_1,\mathbf{r}_2)\Phi^+(\mathbf{r}_4,\mathbf{r}_3)\Phi(\mathbf{r}_3,\mathbf{r}_4)d\mathbf{r}_3 d\mathbf{r}_4 \qquad (2)$$

$$+ \int R_{ex}(\mathbf{r}_1,\mathbf{r}_2,\mathbf{r}_3,\mathbf{r}_4)\Phi(\mathbf{r}_1,\mathbf{r}_4)\Phi^+(\mathbf{r}_4,\mathbf{r}_3)\Phi(\mathbf{r}_3,\mathbf{r}_2)d\mathbf{r}_3 d\mathbf{r}_4 = \mu\,\Phi(\mathbf{r}_1,\mathbf{r}_2).$$

Here the kernels $R_d$ and $R_{ex}$ are caused respectively by direct and exchange interactions between atoms. All electrical characteristics of the gas including its polarization $\mathbf{P}$ can be expressed in terms of the order parameter.

Generally the polarization $\mathbf{P}$ is represented as a multipole expansion. Correct consideration of all terms in this series is simplified if we switch from the polarization $\mathbf{P}$ to the polarization charge $\rho_{pol}$. These quantities are linked with a relation $\rho_{pol}(\mathbf{r}) = -\operatorname{div}\mathbf{P}(\mathbf{r})$. The polarization charge density at an arbitrary point $\mathbf{r}$ equals to

$$\rho_{pol}(\mathbf{r}) = e\left\langle \Phi_0 \left| \left[ \hat{\Psi}_c^+(\mathbf{r})\hat{\Psi}_c(\mathbf{r}) - \hat{\Psi}_e^+(\mathbf{r})\hat{\Psi}_e(\mathbf{r}) \right] \right| \Phi_0 \right\rangle. \qquad (3)$$

Calculations show that the density $\rho$ is an infinite series with respect to powers of $\Phi(\mathbf{r}_1,\mathbf{r}_2)$ and $\Phi^*(\mathbf{r}_1,\mathbf{r}_2)$. In the low density limit, when the size of an atom is small compared to the average interatomic distance, (3) leads to

$$\rho_{pol}(\mathbf{r}) = e\int d\mathbf{r}' \left[ |\Phi(\mathbf{r},\mathbf{r}')|^2 - |\Phi(\mathbf{r}',\mathbf{r})|^2 \right]. \qquad (4)$$

This expression has a rather transparent physical meaning, and it could be written down based on physical considerations without any calculations. For example, from the definition of $|\Phi(\mathbf{r}',\mathbf{r})|^2$ as the probability of finding the electron at the point $\mathbf{r}$ when the core is at $\mathbf{r}'$, it follows that the probability of finding the electron at $\mathbf{r}$ with an arbitrary core position is obtained integrating $|\Phi(\mathbf{r}',\mathbf{r})|^2$ by the core coordinate $\mathbf{r}'$. Similarly, the probability of finding the atomic core at $\mathbf{r}$ is obtained by integrating $|\Phi(\mathbf{r},\mathbf{r}')|^2$ with respect to the electron coordinate.

Solving the equation for the order parameter, we take into account that the binding energy of the valence electron and the atomic core $\varepsilon_0$ is much greater than the energy of interaction between the atoms $\varepsilon_{int} = gn$ (where $g$ is the coupling constant) and the energy caused by inhomogeneity $\varepsilon_{inhom} = \hbar^2/2ML^2$, where $M$ is the atom mass, $L$ is the characteristic scale of the inhomogeneity. Hence, the solution for $\Phi(\mathbf{r}_1,\mathbf{r}_2)$ can be found as

$$\Phi(\mathbf{r}_1,\mathbf{r}_2) = \psi(\mathbf{R}_{12}) \cdot \left[ \phi^{(0)}(r_{12}) + \phi^{(1)}(r_{12},\mathbf{R}_{12}) \right]. \qquad (5)$$

Here we replaced the electron coordinate $\mathbf{r}_1$ and the core one $\mathbf{r}_2$ with the coordinate of the center of mass $\mathbf{R}_{12}$ and the relative coordinate $\mathbf{r}_{12}$. The function $\phi^{(0)}$ is the ground state wave function of the atom, and $\psi(\mathbf{R}_{12})$ must satisfy the Gross-Pitaevskii equation which arises in this theory as the solvability condition for the first order (with respect to the interaction between the atoms) equation for $\phi^{(1)}$. In the zero order approximation $\Phi(\mathbf{r}_1,\mathbf{r}_2) = \psi(\mathbf{R}_{12})\phi^{(0)}(r_{12})$. Substituting this $\Phi(\mathbf{r}_1,\mathbf{r}_2)$ into the expression for the charge density $\rho_{pol}(\mathbf{r})$ we obtain, for example, for the first term in (4)

$$\int d\mathbf{r}' |\Phi(\mathbf{r},\mathbf{r}')|^2 = \int d\mathbf{r}'\, \phi_0^2(|\mathbf{r}-\mathbf{r}'|) \left|\psi\left(\frac{m_c\mathbf{r}+m_e\mathbf{r}'}{M}\right)\right|^2 = $$

$$= \int d\mathbf{r}''\, \phi_0^2(r'') \left|\psi\left(\mathbf{r}+\frac{m_e}{M}\mathbf{r}''\right)\right|^2. \qquad (6)$$

Here $M = m_c + m_e$. Noticing that the characteristic length at which $\psi(\mathbf{r})$ varies is much greater than the length at which $\phi_0(r)$ varies, we expand $\psi$ near the point $\mathbf{r}'' = 0$. Taking into account $|\psi(\mathbf{r})|^2 = n(\mathbf{r})$ and keeping only the first nonvanishing term of the expansion, we find from (5)

$$\int d\mathbf{r}'' \phi_0^2(r'') n\left(\mathbf{r} + \frac{m_e}{M}\mathbf{r}''\right) = \int d\mathbf{r}'' \phi_0^2(r'') \left[ n(\mathbf{r}) + \frac{m_e}{M}(\mathbf{r}''\nabla) n(\mathbf{r}) + \frac{1}{2}\left(\frac{m_e}{M}(\mathbf{r}''\nabla)\right)^2 n(\mathbf{r}) + ...\right]. \quad (7)$$

Note that the linear in $\mathbf{r}''$ term turns into zero after integration due to evenness of the function $\phi_0(r)$. In advance we point that $\phi^{(1)}(r)$ has an addition odd with respect to $\mathbf{r}$. As the result, in the polarization charge density in the first order approximation the first nonvanishing term will be linear in $\mathbf{r}''$.

Similar calculations for the second term in (4) show that in the coefficient at the third term in the right-hand side of (7) the electron mass $m_e$ must be replaced with the core mass $m_c$. Finally we find the polarization charge from (4)

$$\rho_{pol}^{(0)}(\mathbf{r}) = \frac{e}{2}\frac{m_e - m_c}{m_e + m_c}\int d\mathbf{r}'' \phi_0^2(r'')(\mathbf{r}''\nabla)^2 n(\mathbf{r}) = -C_0 a_B^2 e\gamma \cdot \nabla^2 n(\mathbf{r}), \quad (8)$$

where $\gamma = (m_c - m_e)/(m_c + m_e)$. Since in any atom $m_c \gg m_e$, it is assumed hereafter $\gamma = 1$. The constant $C_0$ in (8) is of the order of unity. Expression (8) is valid only at large (compared to $a_B$) distances from the system boundary.

Appearance of a macroscopic charge in the system of noninteracting particles seems quite strange. The charge of each atom is zero – why does the macroscopic density $\rho_{pol}(\mathbf{r})$ arise in the spatially inhomogeneous system? This is because, due to the strong inequality $m_c \gg m_e$, the probability to find the atomic core (positive charge) at a certain point $\mathbf{r}_0$ can be assumed proportional to $\delta(\mathbf{r} - \mathbf{r}_0)$, while the negative charge density is determined by the distribution of the valence electron. Therefore the total charge density at each point inside the atom is nonzero. For noninteracting atoms, when there is no preferential direction in the system, $\rho_{pol}$ must be proportional to $\nabla^2 n$.

To find the first order correction $\phi^{(1)}$ in (5), the equation for $\Phi(\mathbf{r}_1, \mathbf{r}_2)$ was solved using the perturbation theory with assumption that the interaction between atoms is small due to diluteness of the system [9]. The correction $\phi^{(1)}$ leads to an additional term in the polarization charge

$$\rho_{pol}^{(1)}(\mathbf{r}) = -C_1 e a_B^2 \nabla \left[ n(\mathbf{r}) \nabla n(\mathbf{r}) \right], \quad (9)$$

where $C_1 \sim 1$ is a dimensionless constant.

Combining the results (8) and (9), we obtain an expression for the electric polarization of the system of weakly interacting atoms

$$\mathbf{P} = \left( C_0 e a_B^2 + C_1 n e a_B^5 \right) \nabla n \equiv A \nabla n. \quad (10)$$

It is important to note that for dilute systems (for which this theory is valid) $a_B^3 n \ll 1$. Therefore the second term in the brackets is much less than the first one and it might seem to be omitted. However, this particular term is responsible for appearance of electric fields outside an inhomogeneous superfluid system.

The electric field at a point $\mathbf{r}_0$ outside the system is determined by the equation

$$\mathbf{E}(\mathbf{r}_0) = -\frac{\partial}{\partial \mathbf{r}_0} \int \frac{\rho_{pol}(\mathbf{r}')}{|\mathbf{r}_0 - \mathbf{r}'|} d^3 r' = -\frac{\partial}{\partial \mathbf{r}_0} \int \frac{e\left(\Phi^2(\mathbf{r}', \mathbf{r}'') - \Phi^2(\mathbf{r}'', \mathbf{r}')\right)}{|\mathbf{r}_0 - \mathbf{r}'|} d^3 r' d^3 r''. \quad (11)$$

Neglecting edge effects, we can swap the integration variables $\mathbf{r}' \leftrightarrow \mathbf{r}''$. Then

$$\mathbf{E}(\mathbf{r}_0) = -\frac{\partial}{\partial \mathbf{r}_0} \int e \Phi^2(\mathbf{r}',\mathbf{r}'') \left( \frac{1}{|\mathbf{r}_0 - \mathbf{r}'|} - \frac{1}{|\mathbf{r}_0 - \mathbf{r}''|} \right) d^3r' d^3r''. \qquad (12)$$

In the case of a system of noninteracting atoms $\Phi(\mathbf{r}',\mathbf{r}'') = \psi(\mathbf{R})\phi_0(r)$. The function $\phi_0(r)$ varies at distances of the order of $a_B$, and the function $\psi(\mathbf{R})$, which describes the motion of an atom as a whole, changes slowly at such distances. This allows to expand the expression in the brackets in (12) with respect to the small ratio $|\mathbf{r}' - \mathbf{r}''|/r$. After changing the variables from $\mathbf{r}'$ and $\mathbf{r}''$ to the center of mass coordinate $\mathbf{R}$ and the relative coordinate $\mathbf{r}$ and integrating by $\mathbf{r}$ we obtain

$$\mathbf{E}(\mathbf{r}_0) = C_1 e a_B^2 \frac{\partial}{\partial \mathbf{r}_0} \int \psi^2(\mathbf{R}) \nabla^2 \frac{1}{|\mathbf{R} - \mathbf{r}_0|} d^3R. \qquad (13)$$

Analyzing this expression, we take into account that $\psi^2(\mathbf{R})$ is the particle density at the point $\mathbf{R}$ and that $\nabla^2(1/|\mathbf{R} - \mathbf{r}_0|) = -4\pi\delta(\mathbf{R} - \mathbf{r}_0)$. Then it follows from (13) that in the absence of interaction between the atoms an inhomogeneity in the system of atoms occupying a certain volume $V$ does not create any electric fields outside this volume, because $\delta(\mathbf{R} - \mathbf{r}_0) = 0$ if $\mathbf{R}$ is inside and $\mathbf{r}_0$ is outside $V$. This result is quite natural since in the absence of interaction inhomogeneously distributed atoms do not create stationary electric fields outside the atoms.

In the presence of interaction the order parameter $\Phi(\mathbf{r}',\mathbf{r}'')$ obtains an addition proportional to $\phi_1(\mathbf{r},\mathbf{R})$, the expression for which has been found in [6, 9]. After substituting to (12) the corresponding expression for $\Phi(\mathbf{r}',\mathbf{r}'')$, we find with linear accuracy in the interaction

$$\mathbf{E}(\mathbf{r}_0) = -C_1 e n_0 a_B^5 \frac{\partial}{\partial \mathbf{r}_0} \int \psi^2(\mathbf{R}) \frac{\partial \psi^2(\mathbf{R})}{\partial \mathbf{R}} \cdot \nabla \frac{1}{|\mathbf{R} - \mathbf{r}_0|} d^3R \equiv -\frac{\partial}{\partial \mathbf{r}_0} \int \mathbf{P}(\mathbf{R}) \cdot \nabla \frac{1}{|\mathbf{R} - \mathbf{r}_0|} d^3R. \qquad (14)$$

The polarization vector $\mathbf{P}(\mathbf{R})$ here coincides with $\mathbf{P}(\mathbf{R})$ in (9) where we should set $C_0 = 0$.

In order to find the electric field (14) it is needed to find the density of the system $n(\mathbf{r}) = |\psi(\mathbf{r})|^2$, where the function $\psi(\mathbf{r})$ is the solution of the Gross-Pitaevskii equation. Writing down $\psi(\mathbf{r},t)$ as $\psi(\mathbf{r},t) = \sqrt{n(\mathbf{r},t)} \exp[i\varphi(\mathbf{r},t)]$, substituting it to the Gross-Pitaevskii equation and separating the real and imaginary parts, we obtain two equations. One of them is the continuity equation, and the other one is analogous to the classical Euler equation

$$\hbar \frac{\partial \varphi}{\partial t} = \frac{\hbar^2}{2M} \frac{\nabla^2 \sqrt{n}}{\sqrt{n}} - \frac{Mv^2}{2} - gn + \mu \qquad (15)$$

where $\mathbf{v} = \hbar \nabla \varphi / M$ is the superfluid velocity.

Let us consider the case of a rectilinear vortex line parallel to the $z$ axis. In the presence of superfluid flows passing the vortex it can move with a velocity $\mathbf{v}_L$. At small velocities $\mathbf{v}_L$ the vortex line moves as a whole without deformation of its structure and

$$\phi(\mathbf{r},t) = \phi(\mathbf{r} - \mathbf{v}_L t). \qquad (16)$$

Hence,

$$\frac{\hbar}{M} \frac{\partial}{\partial t} \phi = -\mathbf{v} \cdot \mathbf{v}_L. \qquad (17)$$

Significant deformation of the density $n$ is known to take place in the vortex core. The radius of this core is $\xi = \hbar(2Mgn_0)^{-1/2}$. In superfluid $^4$He the vortex thickness is of the order of interatomic distance. In a weakly nonideal Bose gas the core can have a macroscopic thickness, but it is always much less than the characteristic size of the system. At distances large compared to the core size we can omit the first term in the right-hand side of (15) containing the coordinate derivatives (Thomas-Fermi approximation) and obtain the following expression for an addition $n'(\mathbf{r},t) = n(\mathbf{r},t) - n_0$ caused by gas motion

$$n' = -\frac{M}{2g} \cdot \left[ \mathbf{v}^2 - 2\mathbf{v} \cdot \mathbf{v}_L \right]. \tag{18}$$

In the case when a uniform superfluid flow with a velocity $\mathbf{v}_{S0}$ is present, the velocity $\mathbf{v}$ in (18) consists of the velocity $\mathbf{v}_v$ of circular motion around the vortex core and of $\mathbf{v}_{S0}$. It follows from (18) and from (10) (with $C_0 = 0$) that the polarization of the gas caused by a rectilinear vortex equals to

$$\mathbf{P} = -\frac{C_1 n_0 a_B^5 M}{2g} \nabla \left[ \mathbf{v}_v^2 + 2\mathbf{v}_v \cdot (\mathbf{v}_{S0} - \mathbf{v}_L) \right]. \tag{19}$$

Expression (19) remains indeterminate while the velocity of the vortex $\mathbf{v}_L$ is unknown. To find it we must go beyond the scope of the microscopic calculation used here which is valid only at $T = 0$. At nonzero temperatures it is necessary to take into account the presence of excitations in the system. In a phenomenological approach the velocity $\mathbf{v}_L$ can be found from the balance equation of forces acting upon the vortex (see e. g. [11…15])

$$\rho_S \kappa \left[ (\mathbf{v}_L - \mathbf{v}_{S0}) \times \hat{\mathbf{z}} \right] = -D(\mathbf{v}_L - \mathbf{v}_n) - \frac{\kappa}{|\kappa|} D' \left[ \hat{\mathbf{z}} \times (\mathbf{v}_L - \mathbf{v}_n) \right], \tag{20}$$

where $\kappa$ is the vortex circulation. The left-hand side of this equation is the Magnus force, the first term in the right-hand side is the friction force and the second one is the Iordanskii force. Coefficients $D$ and $D'$ are determined by scattering of phonons and rotons on the vortex. Solving the balance equation, we obtain the velocity of the vortex line

$$\mathbf{v}_L = \frac{\kappa \rho_S D_1}{T} \left[ \hat{\mathbf{z}} \times \mathbf{w} \right] + D_2 \mathbf{w} + \mathbf{v}_{S0}, \tag{21}$$

where $\mathbf{w} = \mathbf{v}_n - \mathbf{v}_{S0}$ is the relative velocity of superfluid and normal components. Coefficients $D_1$ and $D_2$ are determined by expressions

$$D_1 = \frac{TD}{\left[ |\kappa|\rho_S - D' \right]^2 + D^2}, \quad D_2 = 1 - \frac{|\kappa|\rho_S (\kappa\rho_S - D')}{\left[ |\kappa|\rho_S - D' \right]^2 + D^2}. \tag{22}$$

Now let us find the electric potential above an infinite film of thickness $h$ at a large distance compared to $h$. Using (14), (19), (21) we obtain

$$\phi(\mathbf{r}_0) = \frac{\mathbf{d} \cdot (\mathbf{r}_0 - \mathbf{r}_v)}{|\mathbf{r}_0 - \mathbf{r}_v|^3}. \tag{23}$$

Here

$$\mathbf{d} = \frac{C_1 e n_0 a_B^5 M h}{g} \left( \frac{\rho_S \kappa^2 D_1}{T} \hat{\mathbf{z}} \times \left[ \hat{\mathbf{z}} \times \mathbf{w} \right] + \kappa D_2 \left[ \hat{\mathbf{z}} \times \mathbf{w} \right] \right). \tag{24}$$

The expression for $\mathbf{d}$ can be interpreted as the dipole moment of the vortex. We see that relative motion of normal and superfluid components causes appearance of a dipole moment of the vortex. This dipole moment consists of two parts, one of them is parallel and the other one is orthogonal to the relative velocity $\mathbf{v}$. Note that only the second part depends on the sign of the vortex circulation.

It is useful to estimate the order of magnitude of this dipole moment for helium. To estimate the coupling constant $g = 4\pi\hbar^2 a / M$ we shall assume the scattering length $a \sim 3 \cdot 10^{-8}$ cm. At a temperature of the order of 1 K ($\kappa D_1 \sim D_2 \sim 1$) and at relative velocity $w = 1$ cm/s we obtain an estimate for the dipole moment per unit length $d/h \sim e \cdot 10^{-7}$. Since the dipole moment $\mathbf{d}$ is created by atoms in a cylindrical tube with radius of the order of $\xi$ and length $h$, at $h = 1$ the number of such atoms is $\pi\xi^2 \cdot 1 \cdot \rho_S / M \sim 10^7$ and, therefore, each atom in the vortex has a dipole moment $\sim e \cdot (10^{-14}$ cm$)$. This dipole moment would appear in a helium atom if one places it into an external electric field $E \sim 10^2$ V/cm (polarizability of helium is $\alpha = 2 \cdot 10^{-25}$ cm$^3$).

1. A. S. Rybalko, Low Temp. Phys. **30**, 994 (2004).
2. A. S. Rybalko, S. P. Rubets, Low Temp. Phys. **31**, 623 (2005).
3. L.A. Melnikovsky, J. Low Temp. Phys. **148**, 559 (2007).
4. A. M. Kosevich, Low Temp. Phys. **31**, 839 (2005).
5. V. D. Natsik, Low Temp. Phys. **31**, 915 (2005).
6. S. I. Shevchenko, A. S. Rukin, JETP Lett. **90**, 42 (2009).
7. S. I. Shevchenko, A. S. Rukin, Low Temp. Phys. 36, 146 (2010).
8. S. I. Shevchenko, 26th International Conference on Low Temperature Physics (LT26), A0149 (2011).
9. S. I. Shevchenko, A. S. Rukin, Low Temp. Phys. **38**, 905 (2012).
10. L. V. Keldysh, in: Problems of Theoretical Physics, Nauka, Moscow (1972).
11. H. E. Hall, W. F. Vinen, Proc. R. Sc. London Ser. A **238**, 215 (1956).
12. E. M. Lifshitz, L. P. Pitaevskii, Zh. Eksp. Teor. Fiz. 33, 535 (1957) [Sov. Phys. JETP 6, 418 (1958)].
13. L. P. Pitaevskii, Zh. Eksp. Teor. Fiz. 35, 1271 (1958) [Sov. Phys. JETP 8, 888 (1959)].
14. S. V. Iordanskii, Zh. Eksp. Teor. Fiz. 49, 225 (1965) [Sov. Phys. JETP 22, 160 (1966)].
15. E. B. Sonin, Phys. Rev. B 55, 485 (1997).